\documentclass[twocolumn]{aastex631}
\usepackage{float}
\usepackage{amsmath}
\usepackage{graphicx}
\usepackage{dcolumn}
\usepackage{bm}
\usepackage{threeparttable}
\usepackage{diagbox}
\usepackage{makecell}

\newcommand{\g}{$\gamma$}
\def\bea{\begin{eqnarray}}
\def\eea{\end{eqnarray}}
\def\be{\begin{equation}}
\def\ee{\end{equation}}

\shorttitle{PBH from CMB Lensing and \g-ray Emissions}
\shortauthors{Tan et al.}

\begin{document}
\title{Searching for Signal of Primordial Black Hole from CMB Lensing and \g-ray Emissions}
\author{Xiu-Hui Tan}
\affiliation{Department of Astronomy, Beijing Normal University, Beijing 100875, China}

\author{Yang-Jie Yan}
\affiliation{Department of Astronomy, Beijing Normal University, Beijing 100875, China}

\author{Taotao Qiu}
\affiliation{School of Physics, Huazhong University of Science and Technology, Wuhan, 430074, China}
\author{Jun-Qing Xia}
\affiliation{Department of Astronomy, Beijing Normal University, Beijing 100875, China}
\affiliation{Institute for Frontiers in Astronomy and Astrophysics, Beijing Normal University, Beijing 100875, China}

\begin{abstract}

In this {\it Letter}, we search for the signal of the primordial black holes (PBHs) by correlating the \g-ray emissions in the MeV energy band produced by the Hawking evaporation and the lensing effect of the cosmic microwave background (CMB). We use the conservative case of the astrophysical model as much as possible in the calculations, since the potential astrophysical origins dominate the observed emission in the MeV energy band. By carefully discussing the appropriate energy bands corresponding to different PBHs masses, it is worth expecting a tight constraint on the fraction of the Schwarzschild PBHs in the mass range of $10^{16} - 5\times10^{17}\,{\rm g}$, by simulations of the sensitivity of the future CMB-S4 project and the \g-ray telescope e-ASTROGAM. Furthermore, we also consider the PBHs model with spins, and find that the constraining ability of the PBHs fraction from the correlation between CMB lensing and \g-ray emissions can be improved by another order of magnitude, which could importantly fill the gaps with PBHs fraction limits in the mass range of $5\times 10^{17} - 2\times 10^{18}\,{\rm g}$.

\end{abstract}


\section{Introduction} \label{sec:intro}
Dark matter (DM) as one of the main components of the Universe has been confirmed by numerous astrophysical and cosmological observations \citep{Planck:2018vyg,Pardo:2020epc}. However, there are still many debates on the candidates of DM, which need further investigation \citep{Boveia:2018yeb}. Since the primordial black holes (PBHs) are the rare well-motivated DM model that does not require the physics beyond the Standard Model \citep{Carr:2016drx}, viewing the PBHs as the DM has been reemphasized recently, after the first detection of the binary black hole merger by the LIGO-Virgo collaboration \citep{LIGOScientific:2016aoc}.

Generally, PBHs are formed in the very early period of the cosmic radiation-dominated period, from the Planck time to the post-inflation era \citep{Hawking:1971ei}. There are many mechanisms to its formation in the literature. The main mechanism for generating PBHs is that the original perturbation re-enters the event horizon, becomes the over-density region to collapse, and finally merges to form PBHs. However, regardless of the formation mechanism by which PBHs are, they will radiate by the Hawking evaporation (HE) \citep{Hawking:1974sw}.

Based on the HE theory, PBHs can directly emit various standard model particles, such as photons, neutrinos and electrons/positrons, etc. When these primary particles propagate in the Universe, it is possible to radiate secondary particles through decay or hadronization processes. If the amount of PBHs is sufficient, this emission has the potential to contribute to the observable background radiation.

Due to the relationship between the rate of HE and the mass of PBHs, PBHs with a mass greater than $5\times 10^{14}\,{\rm g}$ can survive until today and contribute as part of DM \citep{MacGibbon:2007yq}. Thanks to their rich phenomenological properties, non-rotational PBHs with masses $M_{\rm PBH}$ between $5\times 10^{14} - 2\times 10^{17}\,{\rm g}$ have been extensively studied currently, and numerous observational data can be used to strictly constrain the key parameter of PBHs, the fraction $f_{\rm PBH} \equiv \Omega_{\rm PBH}/\Omega_{\rm DM}$. However, the limit of $f_{\rm PBH}$ still leaves a wide window for further exploration, for PBHs mass in the range of $2\times 10^{17} - 10^{23}\,{\rm g}$. 
Some recent works on the upcoming AMEGO \citep{AMEGO:2019gny} and e-ASTROGAM \citep{e-ASTROGAM:2017pxr} observational projects studied the PBHs with the mass up to $\sim10^{18}\,{\rm g}$ by searching for the evaporating signature over the Galactic and extra-Galactic \g-ray background \citep{Carr:2009jm,Carr:2020gox,Coogan:2020tuf,Ray:2021mxu,Auffinger:2022dic}. \citep{Ballesteros:2019exr} estimated the sensitivity to the PBH abundance of a future X-ray experiment in the MeV range, with some unresolved astrophysical sources; isotropic X-ray and soft \g-ray background are proved as an extremely promising target by \citep{Iguaz:2021irx}; the Galactic $511$ keV line \citep{DeRocco:2019fjq} and inner Galaxy \citep{Berteaud:2022tws} by INTEGRAL satellite have placed strong constraints on the fraction of dark matter, respectively.

However, we consider the perturbation information of PBHs to perform this analysis differently. In fact, if we regard PBHs as a kind of DM formed by the gravitational interaction, the \g-ray emissions with a certain energy range radiated by the HE can trace the spatial distribution of PBHs in the large-scale structure of the Universe. On the other hand, there is an intrinsic connection between the CMB formed by the primordial perturbation of the early Universe and the large-scale structure of the Universe observed today, therefore we can conclude that there is also a relationship between the \g-ray from the PBHs and the CMB lensing signal. In this {\it Letter}, thereby we use the cross-correlation technique, which has been widely used as a sophisticated DM indirect detection method by many previous works \citep{Ando:2014aoa,Fornengo:2014cya,Xia:2015wka,Regis:2015zka,Cuoco:2015rfa,Branchini:2016glc,Cuoco:2017bpv,Tan:2020fbc,Tan:2019gmb}, to 
search this kind of connection by simulating the future e-ASTROGAM and CMB-S4 \citep{CMB-S4:2016ple} projects and constrain the PBHs fraction.

\section{Window Function} \label{sec:form}
The \g-ray emissions in the MeV range produced by PBHs could trace the density field. Therefore, we have the source field of PBHs, which provides a primordial seed for the specific observable along the line of sight direction, $\textbf{\^{n}}$, $g_{{\rm PBH}}(\chi, \textbf{\^{n}})= \rho_{{\rm PBH}}(\chi, \textbf{\^{n}})$. Inspired by the expression of intensity from \citep{Arbey:2019vqx} and references therein, we could define the window function of PBHs as:
\be
\begin{split}
&W_{\rm PBH}(\chi)=\frac{f_{\rm PBH}\Omega_{\rm DM}\rho_c}{4\pi M_{\rm PBH}}\\
&\times \int_{\delta E}\int_{t_{\rm min}}^{t_{\rm max}} \left[\frac{{\rm d}^2N}{{\rm d}t {\rm d} E_{\gamma}}\right]_{\rm tot} E_{\gamma}(\chi) e^{-\tau [\chi, E_{\gamma}(\chi)] }{\rm d}E_{\gamma}{\rm d}t~,
\end{split}
\label{equ:winpbh2}
\ee
where ${{\rm d}^2N}/{{\rm d}t {\rm d} E_{\gamma}}$ denotes the number of particles $N$ emitted per units of energy and time. $\chi=\chi(z)$ represents the comoving distance, and $\tau[\chi,E_{\gamma}(\chi)]$ is the optical depth, and the time integral runs from $t_{\rm min}=380,000$ years at last scattering of the CMB to $t_{\rm max}={\rm min}(t(M),t_0)$ where $t(M)$ is the PBHs lifetime with mass $M$ and $t_0$ is the age of the Universe.

In the theory of Kerr PBHs, the number density of photons from an evaporating PBH, depending on the PBHs mass $M_{\rm PBH}$ and the dimensionless spin parameter $a_*\equiv J/GM_{\rm PBH}^2\in[0,1]$ ($J$ is the PBHs angular momentum), in a certain energy range and time interval, 
$\left[\frac{{\rm d}^2N}{{\rm d}t {\rm d} E_{\gamma}}\right]_{\rm pri}=\frac{1}{2\pi}\sum_{{\rm dof}}\frac{\Gamma^j}{\exp{\left[E'/T_{\rm PBH}\right]}-(-1)^{2s}}$,
where $\Gamma$ is the grey-body factor for each particle species, $E'=E-\frac{m}{2M}\frac{a_*}{1+\sqrt{1-a_*^2}}$ is the total energy of the particle, $j$ indicates particle species, $s$ refers to the spin of the particle, and $T_{\rm PBH}$ is the temperature from radiation of PBHs. When $a_*=0$, this reverts to the standard Schwarzschild PBHs. If $a_*$ is close to unity, the spin of PBHs will approach the maximum, and the temperature can change by orders of magnitude. 
PBHs formed during radiation time tend to have low spin with $a \lesssim 0.4$ according to a theoretical prediction from \citep{Chiba:2017rvs}. The probability of extreme-spin PBHs are rare according to \citep{Chongchitnan:2021ehn}, thus, we mainly focus on the $a_*=0$ case and give a mild $a_*=0.5$ and an extreme spin $a_*=0.9999$ for comparison.

\begin{figure}[t]
    \centering
    \includegraphics[width=8.5cm]{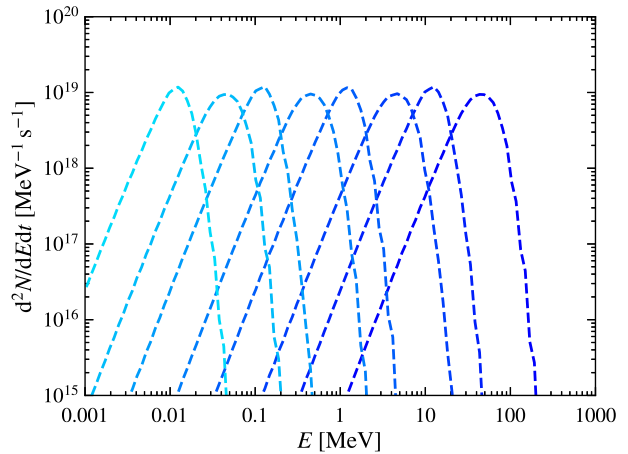}
    \caption{The primary photon spectra ${\rm d}^2N/{\rm d}t{\rm d}E_{\gamma}$ of different $M_{\rm PBH}$ from $10^{15}$ to $5\times10^{18}\,{\rm g}$ with the spin $a_*=0$, calculated by \texttt{BlackHwak}.
    The larger the mass of PBHs is, the lighter the color of the line becomes. From left to right, the lines denote the masses of PBHs: $5\times10^{18}$,  $1\times10^{18}$, $5\times10^{17}$, $1\times10^{17}$, $5\times10^{16}$, $1\times10^{16}$, $5\times10^{15}$, $1\times10^{15}$, respectively.} 
    \label{fig:dndtde}
\end{figure}

The total photon spectrum is combined by the primary component from the direct HE, which is a grey-body factor counting the probability that a Hawking particle evades the PBH gravitational well; and the secondary emission from HE, which is generated from the hadronization and decays.

In our calculations, we use the public software \texttt{BlackHwak} \citep{Arbey:2019mbc,Arbey:2021mbl} to generate the photon spectra with three spin conditions, $a_*=\{0,\,0.5,\,0.9999\}$. In Fig.\ref{fig:dndtde}, we illustrate the primary photon spectra of different mass ($M_{\rm PBH}\in \left[10^{15},5\times10^{18}\right]$ g) for the standard Schwarzschild PBHs, which the maximum of total photon spectra are around the MeV energy band for the PBHs mass range $10^{16} - 5\times 10^{18}\,{\rm g}$. We are only using primary photon spectra to settle down the peak energy of the contribution from PBHs, since the secondary spectra could pull peaks out of the main energy range in our consideration. At the end, we added both of the spectra to estimate the flux.

The CMB lensing convergence, $\kappa$, in a given line of sight is the integral over all the matter fluctuations that will cause gravitational lensing, 
$\kappa(\textbf{\^{n}})=\int{dz W_\kappa(z) \delta(\chi \textbf{\^{n}},z)}$,
where $\delta(\chi \textbf{\^{n}},z)$ is the overdensity of matter at comoving distance $\chi$ and redshift $z$. The distance kernel is given by
\be
W_\kappa(z)=\frac{3}{2}\Omega_{\rm m}H^2_0\frac{1+z}{H(z)}\frac{\chi(z)}{c}\left[\frac{\chi_{\rm cmb}-\chi(z)}{\chi_{\rm cmb}}\right]~,
\ee
where $\Omega_{\rm m}$ is the fraction of the matter density today compared to the present critical density of the Universe, $H_0$ is the Hubble parameter today, $H(z)$ is the Hubble parameter as a function of redshift, $c$ is the speed of light and $\chi_{\rm cmb}$ is the comoving distance to the surface of last scattering where the CMB was emitted. 

As the \g-ray emissions intensity from PBHs is expected to be biased tracers of matter fluctuations, we can use the Limber approximation \citep{Limber:1954zz} to write the cross-correlation power spectrum between \g-ray emissions intensity from PBHs and CMB lensing:
\be
C_\ell^{\kappa,{\rm PBH}}=\int{\frac{{\rm d}z}{c}\frac{H(z)}{\chi^2(z)}W_\kappa(z)W_{\rm PBH}(z)P_{\kappa,\rm PBH}(k=\ell/\chi, z)}~,
\label{equ:clpbh}
\ee
where $P_{\rm \kappa, PBH}$ is the 3D power spectrum of cross-correlation between CMB lensing and the \g-ray emissions from PBHs, which consists of two parts in the halo model, $P_{\rm \kappa, PBH}=P^{\rm 1h}_{\rm \kappa, PBH}+P^{\rm 2h}_{\rm \kappa, PBH}$, and the mass integral runs from $M_{\rm min}=10^7 M_{\odot}$ to $M_{\rm max}=10^{18} M_{\odot}$. For the halo bias of PBHs, we directly take from the equation (12) of \citep{Sheth:1999mn}. Furthermore, we include the contribution from the iso-curvature perturbation to the linear part, as well as the halo mass function of PBHs are calculated, following to \citep{Gong:2017sie}.

\begin{figure}[t]
    \centering
    \includegraphics[width=8.5cm]{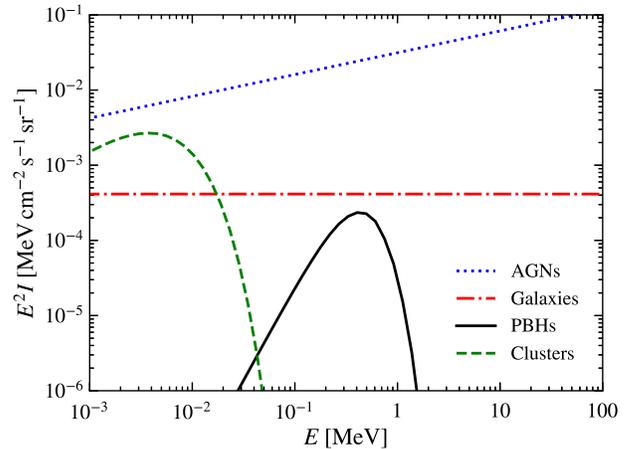}
    \caption{Total intensity produced by AGN (blue dotted line), galaxies (red dash-dotted line), and the cluster of galaxies (green dashed line). For comparison, we also plot the PBHs emission (black solid line) with $M_{\rm PBH}=10^{17}\,{\rm g}$, the spin $a_*=0$ and the fraction $f_{\rm PBH}=10^{-3}$.}
    \label{fig:intensity}
\end{figure}

\section{Astrophysical Sources}
Besides the PBHs emission by the HE, the clusters of galaxies, AGNs and galaxies can also provide the significant contributions to the \g-ray emission of the sky in the MeV energy range. Among these three astrophysical sources, the clusters of galaxies can emit high energy emissions by means of bremsstrahlung radiation of their gas.

We calculated the intensity from bremsstrahlung radiation according to \citep{Zandanel:2015xca} and find that the main contribution of the clusters appears in the energy band $<40$ keV, which is not covered by the energy range of the e-ASTROGAM project. Therefore, in the following calculations, we neglect the contribution from the clusters.

Following \citep{Aird:2009sg,Ptak:2007ae,Caputo:2019djj}, we could obtain the window functions $W(z)$ of AGNs and galaxies:
\be
W_{\rm X}(E,z)=\int^{L_{\rm X,max}(F_{\rm sens},z)}_{L_{\rm X,min}}{\frac{{\rm d}L_{\rm X}}{L_{\rm X}}\Phi_{\rm X}(L_{\rm X},z){\cal L}_{\rm X}(E,z)},
\ee
where $\Phi_{\rm X}(E,z)$ denotes the differential luminosity at an energy $E$ (defined as a number of photons emitted per unit time and per unit energy range, the subscript $\rm X$ refers to AGNs or galaxies) at redshift $z$. We set the minimal luminosity $L_{\rm X,min}=10^{39}\,{\rm erg\,s^{-1}}$, while the maximum luminosity $L_{\rm X,max}$ of an unresolved source is dictated by the sensitivity flux $F_{\rm sens}$ of the \g-ray experiment, providing the minimum detectable flux. The intensity provided by the astrophysical emitters is reported in Fig.\ref{fig:intensity}, where we compare the prediction from the intensity of PBHs.

\begin{figure}[t]
    \centering
    \includegraphics[width=8.5cm]{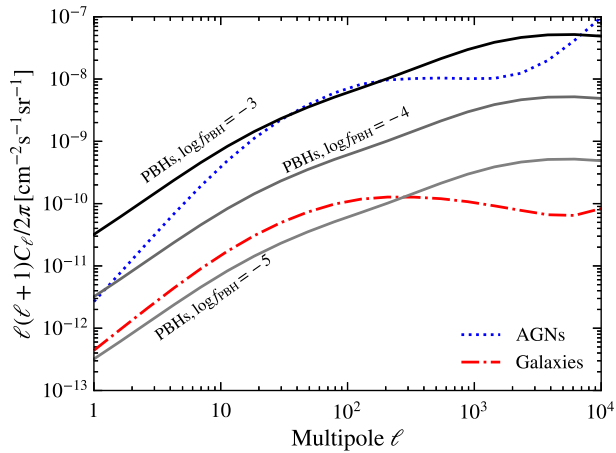}

    \caption{Cross-correlation angular power spectra between the CMB lensing and AGNs (blue dotted line), galaxies (red dash-dotted line), and PBHs (black solid lines). Here, we choose the PBHs mass $M_{\rm PBH}=10^{17}\,{\rm g}$, the spin $a_*=0$. From top to bottom, we show power spectra from different values of the PBHs fraction: $f_{\rm PBH}=10^{-3},\,10^{-4},\,10^{-5}$, respectively.
    } 
    \label{fig:caps}
\end{figure}

Similar with Eq.(\ref{equ:clpbh}), we can also obtain the cross-correlation signals between the CMB lensing and two astrophysical sources, which are shown in Fig.\ref{fig:caps}. 

Obviously, the power spectrum of AGNs contributes most of the signal, while the signal from galaxies is much smaller and mainly on large scales. Here, we also illustrate the signals from PBHs with different fraction values to demonstrate their effects on the cross-correlation power spectrum signal. From the black curves in the figure, we can see that the amplitude of the power spectrum signal is highly dependent on the PBHs fraction, since numerous PBHs will significantly enhance the cross-correlation signal between the CMB lensing and PBHs. The signal from PBHS will be comparable to that from AGN when $f_{\rm PBH}=10^{-3}$.

\subsection{Future Experiments}\label{sec:data}
The e-ASTROGAM observatory \citep{e-ASTROGAM:2018jlu} is dedicated to the study of the non-thermal Universe in the photon range from $0.15$ MeV to $3$ GeV by carrying a \g-ray telescope. 
In this work, we estimate the capability of an instrument with the sensitivity of the e-ASTROGAM to explore the ability of constraints on the PBHs fraction. We adopted the effective area, angular resolution and flux sensitivity varied in energy intervals according to Table III and IV of \citep{e-ASTROGAM:2016bph}, the angular selection is shown in Table.\ref{tab:energy_select} for different mass of PBHs. We take the observational time of $10^6$ s and focus on a fraction of sky $f_{\rm sky}=0.23$, corresponding to the field of view $\Omega=2.9$ sr. The flux sensitivity is $1.1\times 10^{-12}\,{\rm erg\,cm^{-1}\,s^{-1}}$ for all energy ranges and the particle background is 1.4 counts ${\rm s^{-1}\,sr^{-1}}$ in calculation.

\begin{table*}[htp]
    \centering
    \caption{The best energy performance $E_{\rm peak}$ (keV) and angular resolution $\sigma_b$ selection used for different masses of PBHs.}

    \begin{tabular}{l|ccccccccccc}
        \toprule\
        {$M_{\rm PBH}$ (g)} & $1\times 10^{16}$ & $3\times 10^{16}$ & $5\times 10^{16}$ & $7\times 10^{16}$ & $1\times 10^{17}$ & $3\times 10^{17}$& $5\times 10^{17}$& $7\times 10^{17}$& $1\times 10^{18}$ & $3\times 10^{18}$ & $5\times 10^{18}$\\
        \hline\\
         $a_*=0$ & 4690& 2234& 1204& 831& 448& 213&150&102&50&23&12\\\\
         $a_*=0.5$ & 4690& 2234& 1362& 940& 507 & 213& 150&102&50&23&12 \\\\
         $a_*=0.9999$ &7690& 3662  & 2234 & 1541 & 831  & 350 & 213&167&79&38&20\\\\
         $\sigma_b$& $0.8^{\circ}$ & $0.8^{\circ}$ & $1.1^{\circ}$ & $1.5^{\circ}$&$1.5^{\circ}$&$2.5^{\circ}$&$4.3^{\circ}$&$4.3^{\circ}$&$4.3^{\circ}$&$4.3^{\circ}$&$4.3^{\circ}$\\\\
         \hline
    \end{tabular}
    \label{tab:energy_select}
\end{table*}

As we show in Fig.~\ref{fig:dndtde}, the photon spectra, as a function of the energy band, has the peak at the different energy for different mass of PBHs. Correspondingly, we list the peak energy $E_{\rm peak}$ for different $M_{\rm PBH}$ with three values of spin in Tab.\ref{tab:energy_select}. Apparently, the larger the mass of PBHs is, the lower energy peak it has. We notice that when considering the mass of PBHs $M_{\rm PBH}>5\times10^{17}\,{\rm g}$, the peak energy $E_{\rm peak}$ has already out of the energy range of e-ASTROGAM. Therefore, For these masses of PBH, we did not consider photons near the peak energy (see Table.\ref{tab:energy_select}), but only photons within the lowest energy band of the energy range of the e-ASTROGAM, i.e., from 150 keV to 300 keV.

For CMB lensing, we assume a CMB-S4 experiment \citep{CMB-S4:2016ple} with the telescope beam of Full-Width-Half-Maximum (FWHM) of $1'$ and a white noise level of $1~\mu K'$ for temperature and $1.4~\mu K'$ for polarization. We set the noise levels $N_\ell^{\rm TT}$ and $N_\ell^{\rm EE}$ in the primary CMB as a Gaussian noise as:
$N_\ell^{\rm XX} = s^2{\rm exp}\left(\ell(\ell+1)\frac{\theta^2_{\rm FWHM}}{8{\rm log}2}\right)$,
where ${\rm XX}$ stands for ${\rm TT}$ or ${\rm EE}$, here $s$ is the total intensity of instrumental noise in $\mu K\,{\rm rad}$, and $\theta^2_{\rm FWHM}$ is the {\rm FWHM} of the beam in radians. For the CMB lensing reconstruction noise, we use the EB quadratic estimator method described in \citep{Hu:2001kj}, implemented by the \texttt{QUICKLENS} software package \footnote{\url{https://github.com/dhanson/quicklens/}}.

Finally, we compute the theoretical power spectra and perform the numerical constraint on the PBHs fraction. The redshift ranges we consider is from 0 to 10, which is reasonable for both PBHs and the astrophysical sources. Furthermore, we assume the flat $\Lambda$CDM cosmology with parameters set by \textit{Planck} results \citep{Planck:2018vyg}: $h = 0.6766$, $\Omega_{\rm b}h^2=0.02243$, $\Omega_{\rm c}h^2=0.11999$, $\tau=0.0561$, and $n_s=0.9665$.

\section{Constraint Results}
Assuming the experiments are the power spectra of Gaussian random fields, we can compute the covariance matrix as:
\be
\begin{split}
&\Gamma^{\gamma, {\kappa}}_{\ell, \ell'}=
\frac{\delta_{\ell\ell'}}{\left(2\ell+1\right) f_{\rm sky} \Delta \ell} \\
&\times \left\lbrack C^{\gamma {\kappa}}_{\ell}C^{\gamma {\kappa}}_{\ell'} + \left( C_{\mathcal{N}}^\gamma+\sqrt{C_{\ell}^\gamma C_{\ell'}^{\gamma}}\right) \left(C_\mathcal{N}^{\kappa}+C_\ell^{\kappa} \right)  \right\rbrack, \\
\end{split}
\label{equ:cov}
\ee
where the photon noise term above is $C^\gamma_{\mathcal{N}}=4 \pi f_{\rm sky} \left\langle I_{\rm X} \right\rangle^2 N_{\rm X}^{-1} W_\ell^{-2}$, and $\left\langle I_{\rm X} \right\rangle$ refers to the sky-averaged intensity observed by the telescope, and is assumed that comes from the AGNs and galaxies contributions by default; $N_{\rm X}=\left\langle I_{\rm X} A_{\rm eff}\right\rangle t_{\rm obs}\Omega_{\rm FoV}$; the beam window $W_\ell=\exp{\left(-\sigma^2_{\rm b}\ell^2/2\right)}$ is a Gaussian point-spread function, where $\sigma_{\rm b}$ is the angular resolution of the instrument, which is described in section \ref{sec:data} and adopted to the Table III from \citep{e-ASTROGAM:2017pxr}, and Table.\ref{tab:energy_select} gives the values for every mass of PBHs.

The upper limits on the parameter space of the PBHs fraction are derived at 95\% confidence level by requiring $\chi^2=2.71$ with the estimator assumed to follow a $\chi^2$ distribution with one degree of freedom, which is
$\chi^2 =\sum_{\ell}\left(C_{\ell}^{\gamma,{\kappa}} \Gamma^{-1}_{\ell \ell'} C_{\ell'}^{\gamma, {\kappa}}\right)$,
\label{equ:chi2}
where the sum of multipoles is from $\ell_{\rm min}=10$ to $\ell_{\rm max}=500$,  which is corresponding to the angular scale $0.36^{\circ}$. Since the angular resolutions $\sigma_b$ used in Eq.(\ref{equ:cov}) is larger than $0.8^{\circ}$, in practice, we find that the photon shot noise term will dramatically increase at high multipoles. Therefore, this choice of $\ell_{max}$ does not affect the final results basically. It is worthy to note that only the cross-correlation term involving the PBHs is present in the signal part of $\chi^2$ the function, since we assume to be able to extract the background associated to the emitting AGNs and galaxies to a good precision and neglect model uncertainties in the astrophysical components. Furthermore, we do not sum up the  energy bins, since for each mass of PBHs, the energy range for integration will be round the different $E_{\rm peak}$ as we mentioned before.

\begin{figure}[t]
    \centering
    \includegraphics[width=8.5cm]{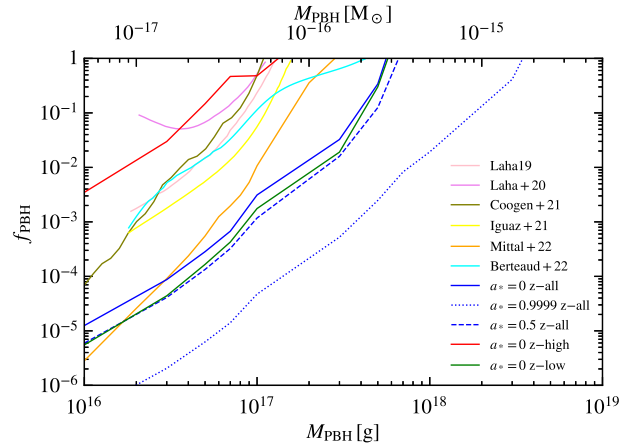}
    \caption{95\% C.L. bounds on the PBHs fraction as a function of the PBHs masses, when considering different cases. See text for details.} 
    \label{fig:constraint_line}
\end{figure}

In Fig.\ref{fig:constraint_line} we present the 95\% C.L. constraints on the PBHs fraction for different $M_{\rm PBH}$. The blue solid line is our main result for the standard Schwarzschild PBHs. The limit on $f_{\rm PBH}$ is around $10^{-5}$ for $M_{\rm PBH}=10^{16}\,{\rm g}$, and down to $10^{-3}$ when increasing the mass of PBHs to $10^{17}\,{\rm g}$. The cross-correlation between \g-ray emissions and CMB lensing significantly compress the allowed space of $f_{\rm PBH}$ for low mass PBHs. 
This constraint is much tighter than the limits on the $f_{\rm PBH}$ from other observations \footnote{\url{ https://github.com/bradkav/PBHbounds/}} by about two orders of magnitude, such as the result based on Comptel observations of the central region of the Galaxy (the olive line) \citep{Coogan:2020tuf}, and as well as the recent limitation from the 21-cm absorption signal of EDGES experiment (the orange line) at $M_{\rm PBH}>10^{17}\,{\rm g}$ \citep{Mittal:2021egv}. 
We also exhibit the outputs from some important PBHs works for comparison, the Galactic Center 511 keV line \citep{Laha:2019ssq} (the pink line); from published Spectrometer aboard the INTEGRAL satellite (SPI) results \citep{Laha:2020ivk} (the purple line); using cosmic X-ray background \citep{Iguaz:2021irx} (the yellow line) and from diffuse soft \g-ray emission towards the inner Galaxy as measured by the SPI data \citep{Berteaud:2022tws} (the cray line).

The reason for this tight constraint is the flux sensitivity of e-ASTROGAM, which is also the key parameter for the photon noise term. Since the low sensitivity, the sky-averaged intensity observed by the telescope $\left\langle I_{\rm X} \right\rangle$ will be significantly larger than that when using other experiments with high sensitivity, like the Athena project \citep{Nandra:2013jka}. This will suppress the shot noise term in the covariance matrix of Eq.(\ref{equ:cov}) and improve the constraints on $f_{\rm PBH}$.

\section{Conclusions \& Discussions}
In the calculations above, we consider the cross-correlation signal in the redshift range $\left[0,\,10\right]$ ($z$-all), which gives very tight constraints on the PBHs fraction for the PBHs mass range $\left[10^{16},\,10^{18}\right]$ g. 
In order to investigate the constraining power on the PBHs fraction further, we split the redshift range in two other ranges for integration: $\left[0,\,1\right]$ ($z$-low) and $\left[1,\,10\right]$ ($z$-high). The reason for separating these three bins is to roughly determine the majority redshift region of PBHs that can be distinguished from astrophysical sources.
In Fig.\ref{fig:constraint_line}, we also show the constraints on $f_{\rm PBH}$ for the $z$-low (green solid line) and $z$-high (red solid line) cases. If we only take the low-redshift information with $z$-low case into account, we obtain that the limitation on $f_{\rm PBH}$ is a slightly tighter to the $z$-all case, which means that the constraining power for PBHs can be more visible in the low-redshift Universe. For the $z$-high case, since we discard the low redshift information, the constraining power on $f_{\rm PBH}$ becomes very weak. If we use the $z$-all, which includes both the low and high redshift information, the constraint on the PBHs fraction becomes an averaged consequence of them.

All of the above conclusions are calculated in the standard Schwarzschild case $a_*=0$. In fact, we also computed the photon spectra with three different spins. When we consider the PBHs with high rotation, the cross-correlation signal with CMB lensing will be significantly enlarged. Consequently, the constraints on the PBHs fraction will be improved in the high-rotating PBHs. In Fig.\ref{fig:constraint_line}, we also showed the limits of $f_{\rm PBH}$ as a function of $M_{\rm PBH}$ for different spins: $a_*=0.5$ (blue dashed line) and $a_*=0.9999$ (blue dotted line). For the $a_*=0.5$ case, the constraints are only slightly enhanced; while in the $a_*=0.9999$ case, the limitation on $f_{\rm PBH}$ becomes significantly improved by more than an order of magnitude. The 95\% C.L. limitation on the PBHs fraction is around $10^{-2}$ at about $M_{\rm PBH}=10^{18}\,{\rm g}$. And the limitation will be close to 1 until the PBHs mass $M_{\rm PBH}$ is larger than $\sim2\times10^{18}\,{\rm g}$. This cross-correlation analysis could importantly fill the gaps with PBHs fraction limits in the mass range $5\times 10^{17} - 2\times 10^{18}\,{\rm g}$. 

Finally, we conclude that based on the future projects, like e-ASTROGAM and CMB-S4, we obtain very tight limitations on the fraction of the Schwarzschild PBHs in the mass range $10^{16} - 5\times10^{17}\,{\rm g}$. The constraining ability of the spin PBHs fraction can be improved by more than one order of magnitude, which could importantly fill gaps of PBHs fraction limits in the mass range $5\times 10^{17} - 2\times 10^{18}\,{\rm g}$.

\section*{Acknowledgments}
We thank Marco Regis and Marco Taoso for useful suggestions on the calculations. J.-Q. Xia is supported by the National Science Foundation of China under grants no. U1931202 and 12021003, and the National Key Research and Development Program of China under grant no. 2020YFC2201603. T. Qiu is supported by the National Science Foundation of China under grants no. 11875141, and the National Key Research and Development Program of China under grant no. 2021YFC2203100.

\bibliography{cmb_pbh}

\end{document}